\documentclass[12pt,preprint]{aastex}

\newcommand{\FeI}{\ion{Fe}{1}}

\def\onedimensional{one-di\-men\-sion\-al}
\def\twodimensional{two-di\-men\-sion\-al}
\def\threedimensional{three-di\-men\-sion\-al}
\def\multidimensional{multi-di\-men\-sion\-al}

\usepackage{mathptmx}
\usepackage{courier}
\usepackage{graphicx}
\usepackage{colordvi}
\usepackage{natbib}
\usepackage{color}

\newcommand{\serenaemail}{serena.criscuoli@oa-roma.inaf.it}
\newcommand{\hanemail}{huitenbroek@nso.edu}

\shortauthors{Uitenbroek, H. \& Criscuoli, S.}
\shorttitle{Comparing one- and \threedimensional\ solar models}

\begin{document}

%
%

\title{Why \onedimensional\ models fail in the diagnosis of average spectra
       from inhomogeneous stellar atmospheres}
\author{Han Uitenbroek}
\affil{{National Solar Observatory/Sacramento Peak}
\footnote{Operated by the %
       Association of Universities for Research in Astronomy, Inc. (AURA), %
       for the National Science Foundation}, P.O.~Box 62, Sunspot, %
       NM 88349, U.S.A.; \hanemail}

\author{Serena Criscuoli}
\affil{INAF -- Osservatorio Astronomico di Roma, %
       Via Frascati 33, 000-40 Monte Porzio Catone, Italy; \serenaemail}

\date{Version \today}

%
%

\begin{abstract}
We investigate the feasibility of representing a structured \multidimensional\
stellar atmosphere with a single \onedimensional\ average stratification
for the purpose of spectral diagnosis of the atmosphere's average spectrum.
In particular we construct four different \onedimensional\ stratifications
from a single snapshot of a magneto-hydrodynamic simulation of solar convection:
one by averaging its properties over surfaces of constant height, and three
different ones by averaging over surfaces of constant optical depth at 500 nm.
Using these models we calculate continuum, 
and atomic and molecular line intensities and their center-to-limb variations.
From analysis of the emerging spectra we identify three main reasons why
these average representations are inadequate for accurate determination
of stellar atmospheric properties through spectroscopic analysis.
These reasons are: non-linearity in the Planck function with temperature,
which raises the average emergent intensity of an inhomogeneous atmosphere
above that of an average-property atmosphere, even if their temperature-optical
depth stratification is identical; non-linearities in molecular formation
with temperature and density, which raise the abundance of molecules of
an inhomogeneous atmosphere over that in a \onedimensional\ model with
the same average properties; the anisotropy of convective motions,
which strongly affects the center-to-limb variation of line-core intensities.
We argue therefore that a one-dimensional atmospheric model that reproduces the
mean spectrum of an inhomogeneous atmosphere necessarily does not reflect
the average physical properties of that atmosphere, and are therefore
inherently unreliable.
\end{abstract}


%
%

\section{Introduction}\label{sec:introduction}
Stellar spectra contain rich information about the physical
conditions of the stars from which they originate, but this
information can only be extracted meaningfully with
sufficiently realistic stellar model atmospheres and
line formation theory.
Traditionally, applicable models have consisted of
\onedimensional\ plane-parallel stratifications
in which the run of temperature with height is determined from 
flux conservation, the mixing length formalism to represent
convective energy transport, and hydrostatic equilibrium.
This homogeneous and static picture, however, is very much at odds with
high-resolution observations of the surface of our closest star, the Sun,
which appears structured even
at the smallest observable spatial scales, and dynamic down to the shortest
temporal scales detectable with current instrumentation,
most particularly at wavelengths that coincide with spectral lines.

Building \onedimensional\ atmospheric representations that account
for this spatial structure and temporal variation is not straightforward,
even if the intent of the model is to reproduce the average spectrum of
such an atmosphere.
First, the model's stratification must represent in some way the
horizontally averaged thermodynamic properties of the
\threedimensional\ atmosphere.
Secondly, it should represent convective motions and resulting Doppler
shifts and their effect on line formation,
and finally, if it is to be physically self-consistent,
it must allow a complex structure supported by
(magneto-)hydrodynamic forces to be represented by hydrostatic equilibrium,
supplemented perhaps by some form of turbulent pressure.
These approximations necessarily involve free parameters, for instance
in the form of a mixing length parameter, micro- and macro turbulent
broadening, and collisional line broadening, to match line
widths, all of which may be freely adjusted to accurately reproduce certain
observables for a given model.
As a result measurement and model are not independent, and this
draws into question the uniqueness of the measurement process.
Finally, adiabatic cooling resulting from rapid expansion in the upper
layers of true convective models is difficult to
represent adequately in \onedimensional\ average models.
In particular, in metal-poor stellar atmospheres this cooling leads to
substantially cooler upper layers than predicted by radiative equilibrium,
whereas in the case of more solar like metallicities the effect is much less pronounced
as the adiabatic cooling is compensated by radiative heating in the multitude
of weak spectral lines \citep{Asplund2005}.
In the upper photosphere of such metal-poor stars the temperatures may be
as much as several hundred to one thousand degrees lower than predicted by
a \onedimensional\ radiative equilibrium model constructed with the same
stellar parameters.

Self-consistent theoretical \onedimensional\ models are not the only ones used for
spectroscopic diagnostics.
In the special case of modeling the solar atmosphere the availability of high
resolution spectra and information on limb darkening, allow for an alternative
to the constraint of flux conservation, namely a semi-empirical determination
of the temperature stratification, with the successful model by
    \citet{HOLMUL}
perhaps the best known example.
Semi-empirical models have no requirement for physical self-consistency,
but as they are constructed to serve a specific set of observables,
the question arises if they can equally reproduce others than those on which they
are based.
    \citet{Holweger+Heise+Kock1990} 
argue, that the answer to this question is affirmative when considering line strengths
used for abundance determinations, because the strengths of different
lines vary in similar fashion, depending mostly on the local temperature
gradient, with lines being stronger in steep gradients, and weaker in
shallow ones, despite differences in excitation and ionization.

To better account for the inhomogeneous and dynamic nature of stellar
atmospheres than possible with \onedimensional\ modeling,
more realistic models have been introduced over the last two decades
that solve the equations for (magneto-)hydrodynamic forces
in a gravitationally stratified atmosphere, consistently with radiative
transfer
    \citep[see][for an overview and references therein]%
{Stein+Nordlund+Asplund-lrsp}, eliminating most of the free
parameters that plague \onedimensional\ atmospheric representations.
Many sophisticated codes now exist for these simulations
    \citep[e.g.,][]{Stein+Nordlund1998,Freytag+Steffen+Dorch2002,%
Schaffenberger_etal2005,Voegler_etal2005,Abbett2007,Jacoutot+Kosovichev+Wray+Mansour2008,%
Hayek_etal2010,Muthsam_etal2010}.
These numerical simulations of solar magneto-convection
have been highly successful in reproducing the morphology of granules
\citep{Stein+Nordlund1989,Stein+Nordlund2000}, the prediction of solar
$p$-modes \citep{Nordlund+Stein2001,Stein+Nordlund2001}, and in
particular in the reproduction of the space- and time-averaged
shapes of photospheric absorption lines
\citep{Asplund+others2000,Asplund+Nordlund+Trampedach+Stein2000}.
They have, however, also stirred a controversy in the determination of
solar abundances, prompting a significant downward revision of
the abundances of oxygen
    \citep{AllendePrieto+Lambert+Asplund2001,%
Asplund+Grevesse+Sauval+AllendePrieto+Kiselman2004}, and carbon
    \citep{AllendePrieto+Lambert+Asplund2002,Asplund+Grevesse+Sauval+AllendePrieto+Blomme2005}
by almost a factor of two, in sharp contradiction
to values determined from \onedimensional\ modeling,
and more importantly, from helioseismology
\citep[see][for a recent overview]{Serenelli+Basu+Ferguson+Asplund2009},
eliciting the question if even these \threedimensional\
convection simulations need further refinement.

With currently available computer resources it is, however, not yet
practical to employ self-consistent \threedimensional\ 
(magneto-) hydrodynamic simulations for all stellar spectroscopic analysis. The
computational task is simply too vast. Ultimately though, many
observations will have to be analyzed in the context of such
modeling, or the validity of much less demanding \onedimensional\
modeling will have to be more firmly established by comparing
with the more realistic \threedimensional\ solutions.
Meanwhile, (semi-empirical) \onedimensional\ models continue to be
employed and are in many instances still at the forefront of stellar
spectroscopic analysis, simply because no applicable self-consistent
models exist yet.
Examples of the latter are analysis of chromospheric spectra
\citep{SocasNavarro+Uitenbroek2004, SocasNavarro2007,%
Avrett+Loeser2008, Centeno+Trujillo+Uitenbroek+Collados2008,%
Grigoryeva+Teplitskaya+Ozhogina2009, Ermolli_etal2010},
irradiance variations resulting from large scale magnetic
phenomena, spectral irradiance in the UV \citep{Fontenla_etal2009,Shapiro_etal2010},
and cases where the curvature of the
atmosphere plays a role, like interpretation of spectra of giants
and supergiants, and spectra taken close to the limb in smaller stars.
 
In an early paper
\citet{Wilson+Williams1972} discuss the effect of inhomogeneities on continuum intensity
of an inhomogeneous atmosphere in the context of modeling the atmosphere of a sunspot umbra.
Several contributions heretofore
\citep{Kiselman+Nordlund1995,Steffen+Ludwig+Freytag1995,Shchukina+TrujilloBueno+Asplund2005,%
Ayres+Keller+Plymate2006,Scott+Asplund_etal2006,%
Pereira+Asplund+Kiselman2009,Ramirez+etal2009,%
TrujilloBueno+Shchukina2009,Caffau+etal2010} have spectroscopically
compared the results from \threedimensional\ simulations with those from
established \onedimensional\ models.
\citet{Koesterke+AllendePrieto+Lambert2008} compared the center-to-limb
variation (CLV) of intensity in continua and lines computed from snapshots of a
\threedimensional\ hydrodynamic simulation to values obtained from a \onedimensional\
model derived by spatially and temporally averaging their snapshots, and
compared their behavior with observed CLVs, and found that the continuum
intensities from their \threedimensional\
snapshots vary stronger with heliocentric angle than those from the
average model (their figure 2).
\citet{Kiselman+Nordlund1995} compared oxygen abundances derived from
two \threedimensional\ HD snapshots with those derived from the \citet{HOLMUL} model
and spatially averaged \onedimensional\ models derived from their 
snapshots. They found that molecular OH lines in all of their \onedimensional\
models were weaker than the spatially averaged ones from their simulation.
\citet{Scott+Asplund_etal2006} compared carbon abundances derived from
CO modeling in \threedimensional\ and average \onedimensional\ models and found that
the latter require slightly higher carbon abundance.

An intermediate approach, between one- and \threedimensional\ modeling was
presented by \citet{Ayres+Keller+Plymate2006} in an attempt to circumvent
the numerical burden of full-blown convection simulations, but still account
for thermal variations in the atmosphere. They employed a so-called 1.5D
transfer model in which properly weighted intensity contributions from
5 \onedimensional\ atmospheres with perturbed temperature stratifications
with respect to an average model are added, such that the CLV of
continuum intensity matches observed behavior.
When comparing the oxygen abundance derived from matching CO line equivalent
widths they found that the single averaged model predicts higher oxygen
abundance than the 5 component model, by about 14\% (their table 4).

In this paper we further investigate the usability of \onedimensional\
atmospheric representations by comparing theoretical spectra of several
different spectral lines and continua and their CLV
behavior for models that have the same
average stratification. Our goal is to identify the reasons why the
spectra differ (or are similar) between the one and \threedimensional\
versions of the same average stratification, so that the nature of 
the errors are better understood when employing the less computationally demanding
\onedimensional\ approach. In this approach we limit ourselves to the simplest case
of \onedimensional\ models directly derived from a \threedimensional\ MHD
snapshot by averaging.
\citet{Atroshchenko+Gadun1994} remark that such an average model is in general, 
not a physically consistent model, as
it does not satisfy vertical pressure equilibrium, nor does it have a
proper equation of state, since dynamical forces that support material
in the vertical direction in the \threedimensional\ simulation are
obviously neglected in the \onedimensional\ average.
Like semi-empirical models, our derived models are thus not physically
self-consistent, although perhaps not to the same degree, as semi-empirical models
at least fulfill hydrostatic equilibrium.

To prove that analysis by one-dimensional modeling is problematic there is
in our view no need
to discuss all possible one-dimensional models that reproduce significant parts of
the average spectrum. That is a search without end because such models are not
sufficiently constrained. 
Instead, with the proof that the one and only one-dimensional model that has, on average,
the same exact thermal stratification as our three-dimensional snapshot
\emph{does not} produce the same average spectrum, we can easily turn our reasoning around
and conclude that any one-dimensional model that does reproduce significant
parts of that spectrum \emph{necessarily} has a different thermal stratification and,
therefore, is bound to fail in the reproduction of other parts of the spectrum.

In Section~\ref{sec:modelatmospheres} we describe and discuss the employed model atmospheres
and spectral synthesis. Resulting spectra are compared in Section~\ref{sec:spectra},
and discussion and conclusions are given in Section~\ref{sec:conclusions}.

\section{Model atmospheres and treatment of radiative transfer\label{sec:modelatmospheres}}
\subsection{Three-dimensional model}
The basis of the various solar atmospheric models that we have
constructed is a \threedimensional\ snapshot from a magneto-hydrodynamic
simulation of solar granulation as described by \citet{Stein+Nordlund1998}.
The original simulation snapshot was interpolated in the vertical dimension to
better represent the surface layers and omit the deeper layers at
large optical depths (below $z = 350$ km) that are irrelevant
for the present investigation.
The interpolated cube measures $253\times253\times64$ grid points and
has a spacing of 23.7 km in the two horizontal dimensions and 13.9 km in
the vertical direction.
Even though the employed snapshot was part of a simulation with non-zero
magnetic field, the average field in the vertical direction being 30 G,
we do not take this field into account in the line transfer calculations,
i.e., we neglect the Zeeman effect, because we are mainly interested
in investigating the possibility and implications of representing
a structured \threedimensional\ model with a \onedimensional\ average
atmosphere by comparing the emergent spectra among each other,
and do not compare calculated spectra with observations. 
Hereafter we will refer to the \threedimensional\ snapshot as MHD30G.

\begin{figure}[htb]
  \epsscale{.70}
  \plotone{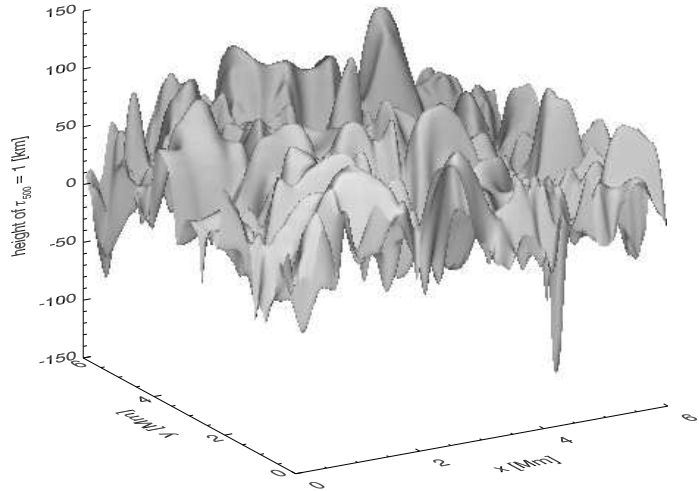}
  \caption{Surface of optical depth unity at 500 nm in snapshot MHD30G.
           \label{fig:ztauone500}}
\end{figure}

The influence of inhomogeneities is illustrated in Figure~\ref{fig:ztauone500},
which shows the surface of optical depth unity at 500 nm in the vertical direction
in MHD30G.
It is clear that surfaces of equal optical depth are
highly corrugated, which causes information that is encoded in the spectrum
at a given wavelength to come from very different geometrical heights as
a function of position on the surface.
The corrugation is the result of the strong temperature sensitivity
of the H$^-$ opacity, which is by far the most dominant absorber
and emitter in the photospheric continuum.
If at a given optical depth the temperature is increased slightly
more free electrons from ionization of metals are produced, the H$^-$
opacity increases and an outward shift of the optical depth scale
takes place, causing the original optical depth to occur at
greater geometric height and lower temperature (since temperature
decreases with height in the photosphere).
This sensitivity causes the contours of equal optical depth to
follow more or less the contours of equal temperature
      \citep{Stein+Nordlund1998}.
Hence temperature fluctuations along surfaces of equal optical
depth are relatively small, much smaller in magnitude than along
surfaces of equal geometrical height, reducing the spatial contrast
in the emergent radiation, compared to what would be expected from
contrast of temperature in surfaces of equal geometrical height.

\subsection{Derived \onedimensional\ models}
\begin{figure}[htb]
  \epsscale{.70}
  \plotone{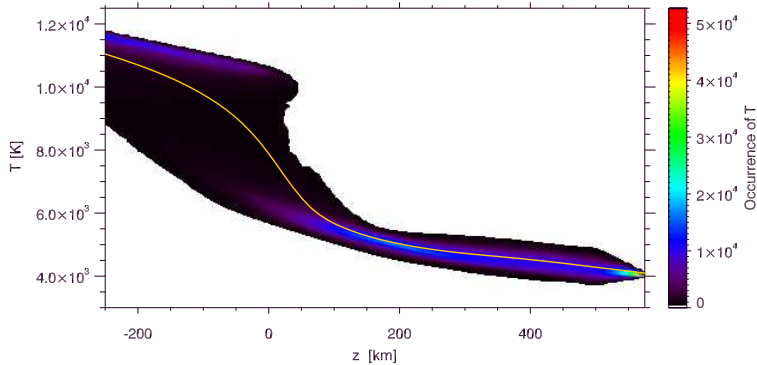}
  \caption{Two-dimensional histogram of temperature values in MHD30G as
           function of geometric height $z$.
           The run of the geometrically averaged temperature is
           plotted with the solid curve.\label{fig:T_histogram_z}}
\end{figure}

We derived four different \onedimensional\ averaged atmospheres from
model MHD30G.
The first model, 1DZ, is derived by averaging all thermodynamical quantities
over horizontal planes.
The temperature stratification of this average model is shown in
Figure\ \ref{fig:T_histogram_z} (solid curve), together
with the two-dimensional histogram of temperatures in MHD30G.
The figure clearly shows the large spread in temperature at any given
height, in particular in the range at or below $z = 0$.
Unfortunately, there is no straightforward way of incorporating the
convective motions present in the \threedimensional\ snapshot realistically
into \onedimensional\ averaged models.
In particular, the horizontal velocity field $(v_x, v_y)$ cannot be directly
represented. Even the average of the vertical velocity $v_z$ cannot be
included if the resulting model is to be static.
Moreover, any vertical velocity would have to be projected into the line of
sight for observing directions off disk center.
Without including any velocities, however, the spectral lines from the
\onedimensional\ model would be much too narrow, and would not sample
similar regions of the atmosphere.
In order to achieve a similar amount of broadening in the spectral lines
calculated from the \onedimensional\ model as in the average spectrum of the
\threedimensional\ snapshot, we chose to include the average velocities
in the commonly employed form of micro-turbulence with a value of one
third of the square root of the sum of the squares of the three
velocity components.
As it turns out, this choice of micro-turbulence closely matches the
photospheric values prescribed in semi-empirical models like the VAL3C model
    \citep{VALIII}
and its later derivatives.
%
%

Since the emergent radiation at a given wavelength typically
comes from a narrow range of optical depths around unity,
the spatially averaged emergent spectrum from an inhomogeneous atmosphere
reflects more the average of atmospheric quantities on an equal optical
depth scale than on an equal geometrical height scale.
A question to ask is therefore if a \onedimensional\ atmosphere that is
constructed to reproduce an average spectrum can accurately reflect
the properties of that atmosphere averaged over equal optical depths.
Of course, even if such a \onedimensional\ representation appears to be
successful for the emergent spectrum in the direction and at the
wavelength it was constructed for, it is unlikely it will be valid for
other directions and wavelengths, since the physical properties
sampled by surfaces of equal optical depth in an inhomogeneous
atmosphere change with direction and wavelength in a much more
complicated way than in a \onedimensional\ atmosphere.
In fact, for more and more inclined viewing angles, a spatially averaged
spectrum gradually changes from representing an average over surfaces
of equal optical depth to one of averages over equal geometrical
height.
\begin{figure}[htb]
  \epsscale{.70}
  \plotone{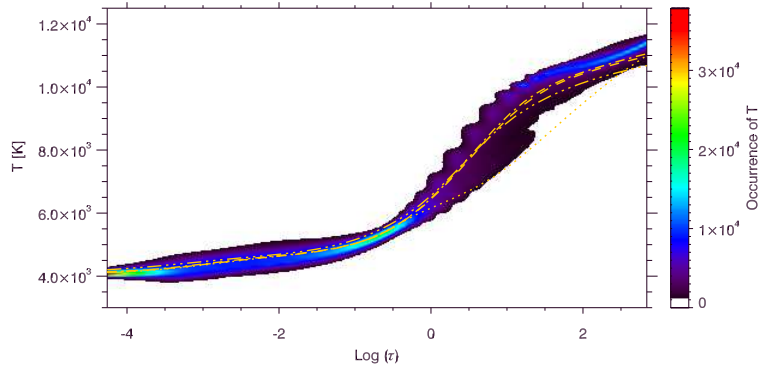}
  \caption{Two-dimensional histogram of temperature in MHD30G as
           function of optical depth at 500 nm in the vertical direction.
           Average temperature stratifications of models 1DTAU (dashed),
           1DT4 (dot-dashed), and 1DHSE (dot-dot-dot-dashed), and
           1DZ (dotted) are plotted against their proper optical depth
           scale at 500 nm.
           \label{fig:T_histogram_tau}}
\end{figure}

To test the premise of how well an equal optical depth averaged model
can reproduce the average spectrum of an inhomogeneous atmosphere and
its CLV we constructed a \onedimensional\
model atmosphere from the MHD30G snapshot by computing the opacities
at 500 nm in each of its columns and reinterpolating all atmospheric
properties to a common optical depth scale at that wavelength 
\citep[e.g.,][]{Kiselman+Nordlund1995,Ayres+Keller+Plymate2006,%
Koesterke+AllendePrieto+Lambert2008}.
We average temperature in two different ways, by averaging $T$ and by averaging
$T^4$ ($\left< T \right> = \left < T^4 \right >^{1/4}$),
as advocated by \citet{Steffen+Ludwig+Freytag1995} to better preserve radiative
flux.
We refer to these models as 1DTAU, and 1DT4 hereafter.
\citet{Steffen+Ludwig+Freytag1995} determined that averaging the fourth power of
temperature to derive the average temperature provided the best way of reproducing
the average spectrum of an inhomogeneous convective model of a white dwarf atmosphere.
Figure~\ref{fig:T_histogram_tau} shows that the differences in stratification
between these two prescriptions of averaging temperature are subtle,
as also found by \citet{Steffen+Ludwig+Freytag1995} in their white dwarf modeling,
and \citet{Steffen+Holweger2002} in \twodimensional\ solar convection modeling
(their figure 3).
A fourth \onedimensional\ model, named 1DHSE, was constructed by taking model 1DTAU
and applying hydrostatic equilibrium, so that the vertical force balance in the
model is self consistent. 
This change of models was accomplished while keeping the relation
between temperature and column mass scale fixed, which preserves the
total mass in the model, but does not preserve the $T$--$\tau_{500}$
relation of the 1DTAU model.  Another option would have been to
transition from one model to the other while keeping the optical depth
scale at a given wavelength (possibly 500 nm) fixed, but we opted for
the former option because it is commonly used in semi-empirical modeling,
and would be the solution favored by a hydrodynamic code programmed to
find a static solution. The latter option, by the choice of the
reference wavelength, introduces even more arbitrariness into the
process of defining one-dimensional models than is already the case
with the choice of the wavelength at which equal opacity surfaces for
averaging are defined.

The obtained average temperature stratifications as a function of the proper optical
depth at 500 nm are drawn in Figure~\ref{fig:T_histogram_tau} over the two-dimensional
histogram of temperature values versus optical depth in the \threedimensional\ snapshot.
Note that the slope of the temperature stratification of model 1DZ is much shallower
than that of both 1DTAU and 1DT4 and to some degree 1DHSE.
The difference arises from the interaction
between the temperature and density structure of the granulation,
and the nature of the H$^-$ opacity, which scales strongly non-linearly
with temperature
     \citep[proportional to $T^{10}$, see][]{Stein+Nordlund1998}.
Because the H$^-$ opacity increases so strongly with temperature,
dense material that resides in the relatively cool
intergranular lanes contributes at higher average temperature
to the geometrically averaged density.
The average opacity of the geometrically averaged atmosphere
is therefore much higher than in the optical depth averaged models,
at the same temperature.
This is visible in the shift towards the right of the dotted
curve in Figure~\ref{fig:T_histogram_tau} with respect to the dashed and dot-dashed
curves, and the dot-dot-dot-dashed curve in the higher layers.
The difference between the two classes of averaged temperature profiles
disappears at smaller optical depths because of the diminished influence
of convection in these layers which results in optical depth surfaces
that are less corrugated in geometrical height.
At much larger depth the topology of the convective flow is dominated
by broad, connected, relatively homogeneous upflows, interspersed
with narrow isolated downflows, again resulting in a convergence of
the $T$---$\tau$ curves, apart from the 1DHSE model.
The latter has a shallower gradient in the deeper layers than 1DTAU and 1DT4,
because the \threedimensional\ stratification is in part supported by
hydrodynamical forces that are not accounted for in hydrostatic equilibrium,
so that the atmosphere has to adjust by increasing pressure and density in
the lower layers, which causes its $T$--$\tau$ curve to shift to the right.

\subsection{Spectral synthesis}
For synthesis of spectra from the different models we used the
one- and \threedimensional\ versions of the transfer code developed by
     \citet{Uitenbroek1998,Uitenbroek2000a,Uitenbroek2001a}.
The different versions of this transfer code share routines that
perform common tasks that are independent of geometry, like the solution of
chemical equilibrium, calculation of line absorption coefficients,
and evaluation of background opacities, which greatly facilitates
comparison between results from the different versions.
The main difference between the one- and \threedimensional\ version
is the way the formal solution of the transfer equation is obtained.
The \onedimensional\ version uses a Feautrier difference scheme as
formulated by
     \citet[][their appendix A]{Rybicki+Hummer1991},
while the \threedimensional\ version employs the short-characteristics
integral method formulated by \citet{Kunasz+Auer1988} as adapted to
periodic boundary conditions by
   \citet{Auer+Fabiani-Bendicho+Trujillo-Bueno1994},
which is much more numerically efficient in \multidimensional\ geometry.
 
\begin{figure}[htb]
  \epsscale{.70}
  \plotone{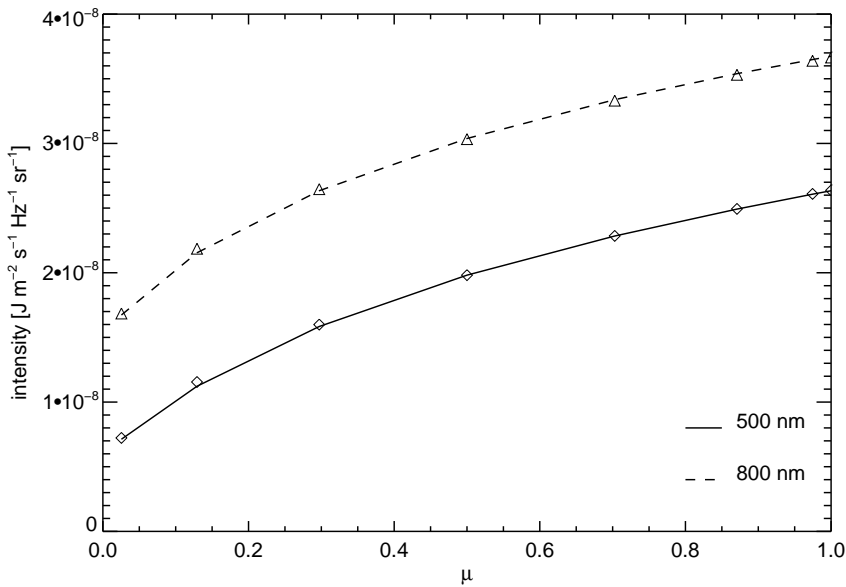}
  \caption{Comparison of CLV of emergent intensities from model 1DZ at 500 nm
           (solid curve) and 800 nm (dashed curve)
           with the spatially averaged intensities (symbols) from a
           \threedimensional\ model
           constructed by replicating the same stratification on
           a $50\times50$ horizontal grid with the same grid spacing
           as in model MHD30G.\label{fig:f1sc3d}}
\end{figure}

To test for possible differences in emergent intensity that could
arise from using two different formal solvers, we created a
\threedimensional\ atmosphere with constant horizontal properties by
replicating our geometrically averaged model 1DZ on a $50\times50$
horizontal grid and calculated the center-to-limb variation of the
emergent intensity at 500 and 800 nm in one- and \threedimensional\
geometry.
The resulting intensities, plotted in Figure\ \ref{fig:f1sc3d}, match
very closely, providing confidence that the two formal solvers
give very similar results when given the same stratification of
source function and opacities, even though they use completely
different numerical methods.

All transfer calculations used in this paper were performed in 
Local Thermodynamical Equilibrium (LTE), except for the contributions
of Thomson scattering by free electrons, Rayleigh scattering by neutral
hydrogen and helium atoms, and H$_2$ molecules, which were lambda iterated.
Other background opacity sources that were included were bound--free and
free--free transitions of H$^-$ and neutral hydrogen, free--free
transitions of H$_2^-$, and H$_2^+$, and bound--free transitions of
different metals.
Instantaneous chemical equilibrium was assumed for all molecular
concentrations and the non-linear set of chemical equilibrium
equations for H$_2$ and CO was solved iteratively with a Newton-Raphson
procedure.

\subsubsection{Wavelengths and spectral lines}
We compare intensities emergent from our three models in several
visible continuum wavelengths, ranging from 400 to 800 nm, one infrared
continuum at 4.7 $\mu$m, two \ion{Fe}{1} lines at 525.0 and 525.3 nm,
respectively, and the 7--6 R69 at 4663.6 nm in the fundamental
rotation-vibration band of the CO molecule.
The continuum wavelengths were chosen to span a range in formation heights
in the solar photosphere corresponding to the maximum of the H$^-$ opacity
(near 800 nm) to its minimum (the H$^-$ opacity at 400 nm is very similar
to that at 1.6 $\mu$m), in the visible range of the spectrum.
In addition, we included the infrared continuum close to the
CO line, also because it represents the spectrum at a wavelength
where the Planck function varies nearly linearly with temperature.

The two iron lines are very close in wavelength and strength, to eliminate
variations resulting from differences in formation height, but have
a significant difference in excitation energy of their lower levels.
The CO line was included to investigate the effect of the non-linear
dependence on temperature of molecular concentrations on line formation.
Parameters for the lines are listed in Table~\ref{tab:lines}.

\begin{table}[htbp]
\caption{Parameters of the spectral lines. \label{tab:lines}}
\begin{tabular}{lcc}
\hline\hline
Line  &  $\log{gf}$  &  $\chi_{\textrm{exc}} [eV]$\\
\hline
\ion{Fe}{1} 525.0209 nm &  -4.938  &  0.121\\
\ion{Fe}{1} 525.3462 nm &  -1.670  &  3.223\\
CO 7-6 R69  4663.6 nm   &  -2.585  &  2.616\\
\hline
\end{tabular}
\end{table}

\subsection{The effect of inhomogeneities on averaging\label{sec:inhomogeneities}}
The horizontal inhomogeneities in temperature and density in the 
atmosphere not only affect the averaging of opacity but also that of
other physical properties that depend on these principal quantities.
Since many of these quantities do not scale linearly with temperature and/or
density, the operation of averaging the properties of an inhomogeneous atmosphere
to a \onedimensional\ stratification (independent of whether this is done
on equal optical depth or geometrical height scales) and evaluating the
sought after physical quantity, does not give the same result as 
evaluating the quantity in the inhomogeneous atmosphere and then averaging
it.

\subsubsection{Molecular densities\label{sec:moldensity}}
An illustrative example for the effect of inhomogeneities in
temperature and density, is the calculation of the number density of a diatomic
molecule AB, which in chemical equilibrium, is given by the Saha equation:
\begin{equation}
  \frac{n_An_B}{n_{AB}}=\left( \frac{2\pi m_{AB} k T}{h^2}\right)^{3/2}
    e^{-D/kT}\left[\frac{U_A(T)U_B(T)}{Q_{AB}(T)}\right] = \Phi(T),
  \label{eq:SahaBoltzmann}
\end{equation}
where $n_{A,B,AB}$ are the number densities for the atomic species A,B
and molecule AB, $m_{AB}$ is the reduced mass of the molecule, $k$
and $h$ are the Boltzmann and Planck constants, respectively, $D$ is
the dissociation energy and $U_{A,B}$ and $Q_{AB}$ are the atomic and
molecular partition functions.
Given Equation\ (\ref{eq:SahaBoltzmann}) it is trivial to see that in general
\begin{equation}
  \left<\frac{n_An_B}{n_{AB}}\right> \neq \Phi(<T>),
\end{equation}
when averaging molecular densities in a region with temperature inhomogeneities.
In particular, non-linear effects are larger for smaller temperatures.
In fact, under the assumptions of low molecular concentrations
($n_{AB}\ll n_A, n_B$) and small temperature perturbations, we can assume that
$n_An_B$ and the partition functions are approximately constant, so that
\begin{eqnarray}
  n_{AB} & = & C \left( \frac{D}{kT} \right)^{3/2} e^{D/kT} \equiv C \phi(T) \label{eq:nAB} \\
  C     & \equiv & n_A n_B \left( \frac{h^2}{2\pi m_{AB} D} \right)^{3/2}
                                                         \left[\frac{Q_{AB}}{U_A U_B} \right]. 
\end{eqnarray}
Let us now estimate the average molecular density $\left< n_{AB}(T) \right>$ in an inhomogeneous
atmosphere with temperature fluctuations $\Delta T$ around $\overline{T} = \left< T \right>$
with $\Delta T / \overline{T} \ll 1$, and compare this average with the concentration
$n_{AB} (\overline{T})$ of the molecule at the average temperature.
Using a Taylor expansion of Equation~\ref{eq:nAB} in $t = \Delta T / \overline{T}$
around $\overline{T}$ we derive to second order in $t$:
\begin{equation}
  \left< \Delta n_{AB} \right> =
    \frac{1}{2} \left< t^2 \right>  C \left.
         \frac{\textrm{d}^2 \phi(t)}{\textrm{d} t^2}\right|_{t = 0} +
         \mathcal{O}(\left< t^3 \right>),
  \label{eq:deltanAB}
\end{equation}
since $\left< \Delta T \right> = 0$ by definition. 
In Appendix~\ref{app:phi} we show that the second derivative of $\phi(t)$ is given by
\begin{equation}
  \phi''|_{t=0} = \left\{\frac{15}{4} + 5 \frac{D}{k \overline{T}} +
                   \left( \frac{D}{k \overline{T}} \right)^2 \right\} n_{AB}(\overline{T}),
\end{equation}
which is strictly positive, so that $< n_{AB} >$ is always greater than
$n_{AB} (\overline{T})$ when temperature fluctuations are present in the
atmosphere.

Note that, for typical temperatures in the solar photosphere of
$\overline{T} = 5.0 \times 10^3$ K
and the dissociation energy of the CO molecule $D_{CO} = 11.091$ eV, the ratio
$D_{CO} / k \overline{T} = 25.7$, causing the Taylor series used for
Equation~\ref{eq:deltanAB} to converge only if typical fluctuations
$\Delta T / \overline{T}$
are smaller than $k \overline{T}/ D_{CO}$, or about 4\%.
In particular for the next term in the expansion to be ignored we should
have $\left< (\Delta T / \overline{T})^3 \right> < (k \overline{T}/ D_{CO})^3$.
However, it is straightforward to show that all even-order derivatives are
positive, so that the claim that the average abundance of a molecule in an
inhomogeneous atmosphere is always larger than the abundance in a homogeneous
atmosphere with the same temperature stratification still holds.

Similarly, it can be shown that density inhomogeneities contribute strictly
positively to the average molecular number density.
In fact, for small molecular concentrations $n_{AB} \ll n_A, n_B$ the
number density $n_{AB}$ scales with the square of the hydrogen density
(see Equation~\ref{eq:SahaBoltzmann}):
\begin{equation}
  n_{AB} \propto A_A A_B\, n_H^2,
\end{equation}
where $A_{A,B}$ are the abundances of elements A and B, and $n_H$ is the hydrogen
number density.
This implies that in an inhomogeneous atmosphere with perturbations $\Delta n_H$
with $\left< \Delta n_H \right> = 0$ the average change in molecular number
density for molecule AB obeys
$\left< \Delta n_{AB} \right> \propto A_A A_B \left< (\Delta n_H)^2 \right >$,
which is again strictly positive for non-zero hydrogen number-density inhomogeneities.

\begin{figure}[htb]
  \plotone{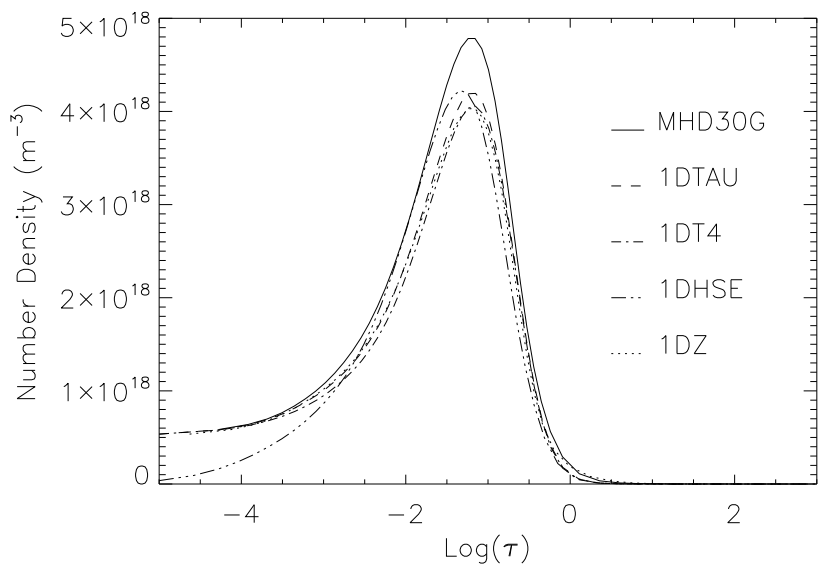}\\
  \caption{CO number densities in snapshot MHD30G averaged over
    surfaces of equal optical depth at 500 nm (solid curve),
    and in the \onedimensional models, 1DTAU (dashed curve)
    1DT4 (dot-dashed), 1DHSE (dot-dot-dot-dashed) and 1DZ (dotted).\label{fig:COnum}}
\end{figure}
The effects of temperature and density inhomogeneities on the
averaging of number densities of the CO molecule is clearly seen in
the difference between the curves in
Figure~\ref{fig:COnum}, which show the average CO number density
over surfaces of equal optical depth of CO in the \threedimensional\
snapshot MHD30G as a function of optical depth (solid curve),
and the run of CO densities in the four different \onedimensional\ models.
Even though the average temperature and hydrogen density in MHD30G are
the same as the temperature and density in the 1DTAU model at each height,
the spatial variation of temperature and density in the \threedimensional\
model causes the average CO density to be higher at each height 
in MHD30G than the CO density at the corresponding height in 1DTAU,
and the same effect is more or less true in the other averaged models
as well.
Interestingly, the CO density at the top of the 1DHSE model drops well
below the concentrations in the other models. This is the result of
enforcing hydrostatic equilibrium. The convective overshoot produces
a high density (and pressure, leading to horizontal expansion) over the
granular surfaces. In establishing hydrostatic equilibrium, which keeps the column
mass -- $T$ relation constant, this over pressure causes
a vertical extension of the atmosphere, reducing densities at the top.
In this reduced density the CO equilibrium shifts towards dissociation.

Since the considered multi-dimensional atmosphere has
both temperature and density inhomogeneities, it is hard to estimate
what the precise contribution from either is to the differences in
average molecular density.
We conclude that average \onedimensional\ models underestimate 
densities compared to the spatially averaged number density in the
inhomogeneous models they are derived from,
and therefore result in an overestimate of the required abundances
of the constituent atoms of the molecules under consideration.
This matches the conclusions by \citet{Kiselman+Nordlund1995} and
\citet{Scott+Asplund_etal2006} who found that oxygen and carbon abundances
derived from molecular diagnostics OH and CO lines, respectively,
were larger for their average \onedimensional\ models than from
their \threedimensional\ snapshots.

\subsubsection{Comparison between low- and high-excitation line}
\begin{figure}[tbhp]
  \plotone{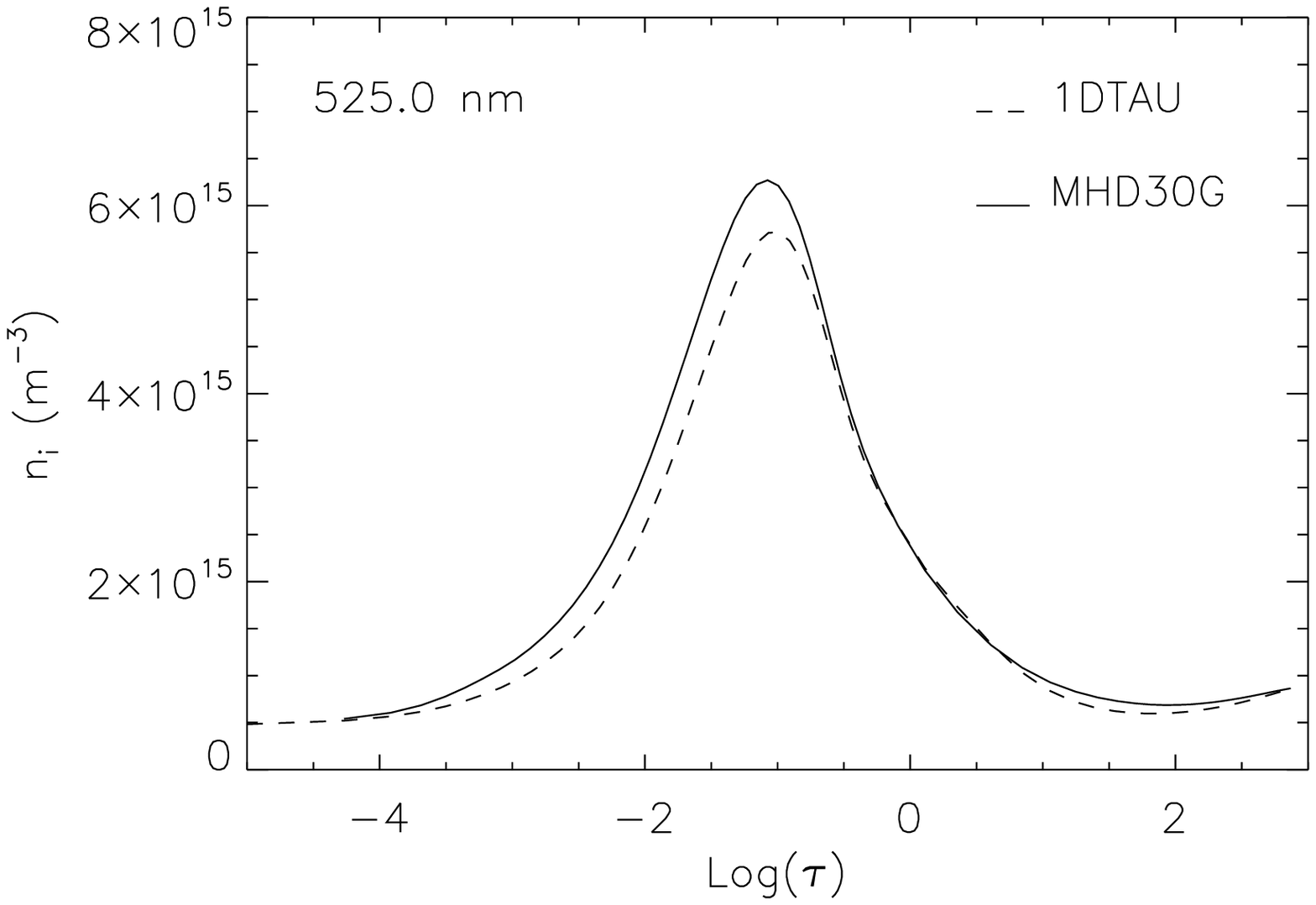}
  \plotone{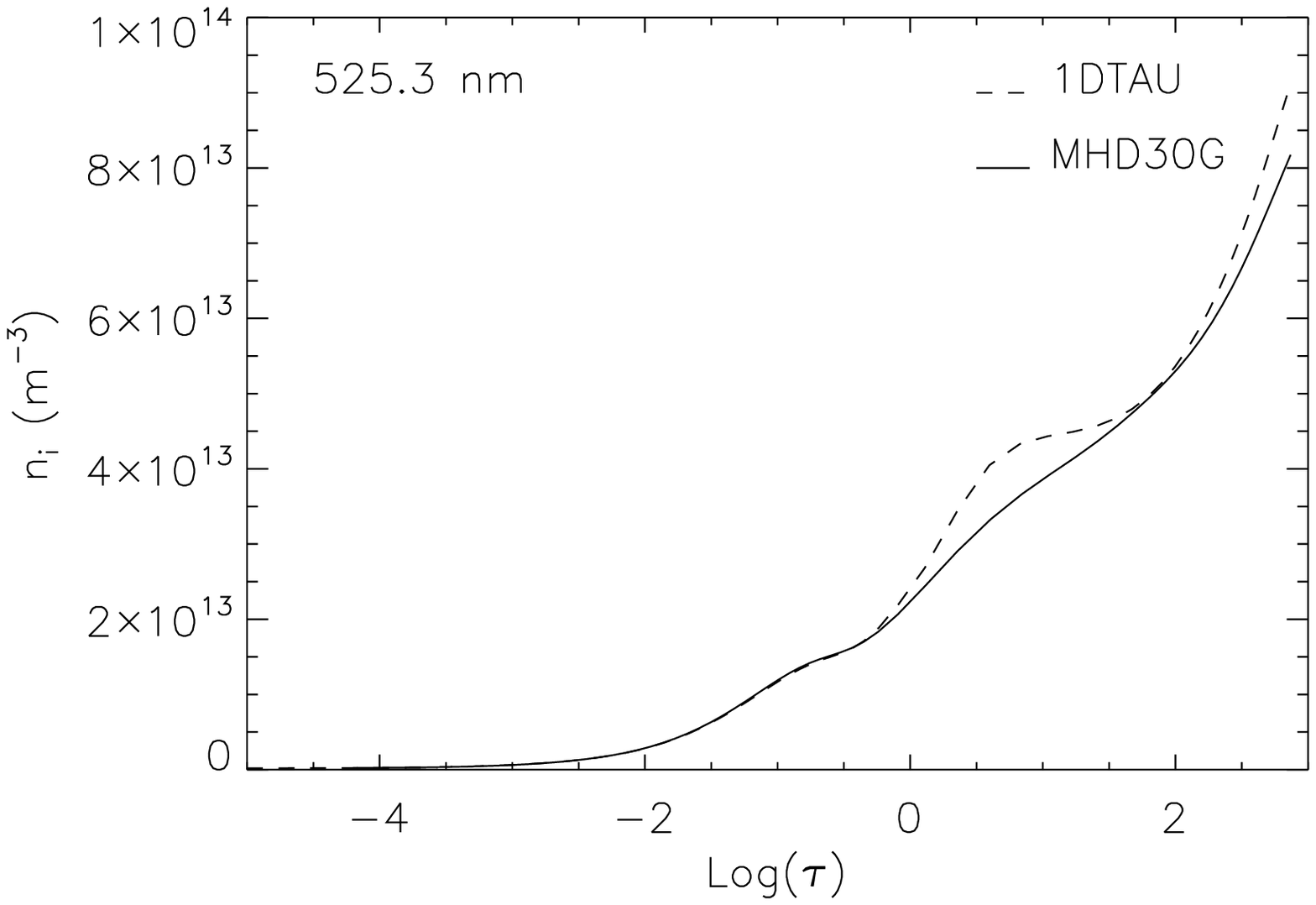}
  \caption{Number densities $n_i$ of the lower level of the low-excitation 525.0 nm
  and high-excitation 525.3 nm lines in MHD30G (solid curve)
  and 1DTAU (dashed).\label{fig:Fenumbers}}
\end{figure}
In our line selection we chose two \ion{Fe}{1} lines of similar strength
and wavelength, but very different excitation potential. Here we
investigate if temperature and density inhomogeneities affect their
lower level populations, and therefore the line opacities, differently.
Indeed, Figure~\ref{fig:Fenumbers} shows that the lower-level populations
are affected very differently: those of the 525.0 nm line are enhanced by the
inhomogeneities, while those of the 525.3 nm line are diminished by them.
The differences in the latter occur below $\tau_{500} = 1$,
and will likely not affect the intensities in the line.
It is also worth to notice that the two lines have similar strength
because the large difference in their excitation potentials,
which is reflected in the large differences of the populations numbers,
is compensated by the difference in their oscillator strength (cf.\ Table~\ref{tab:lines}).

\section{Comparison of spectra\label{sec:spectra}}
\begin{figure}[htb]
\plotone{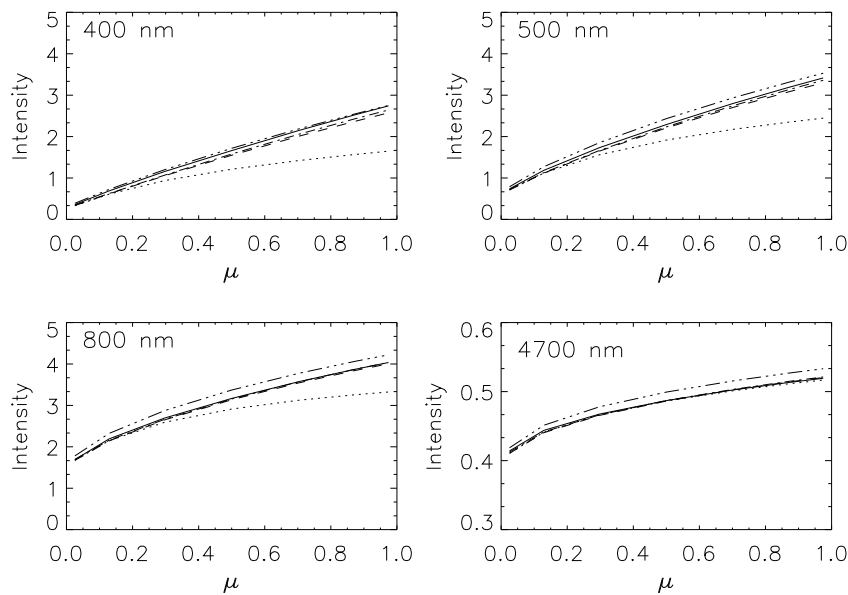}
\caption{CLV of continuum intensities at 400 nm, 500 nm,
  800 nm, and 4700 nm for the four \onedimensional models 1DZ (dotted),
  1DTAU (dashed), 1DT4 (dot-dashed) and 1DHSE (dot-dot-dot-dashed),
  and the spatially averaged continua of MHD30G (solid curves). \label{fig_int_CLV}}
\end{figure}

The emission of radiation is a non-linear process with respect to the
main physical parameters of the atmosphere. As a consequence,
differences must be expected when comparing radiative emission
obtained from the five atmospheric models investigated, even
when the MHD30G and 1DTAU, for instance, have exactly the same average
thermal stratification (in optical depth at 500 nm).
As an example, Figure \ref{fig_int_CLV} shows the CLV
of intensity obtained in four continua, sampling various heights in
the atmosphere, namely 400, 500, 800 nm and 4.7 $\mu$m.
The first thing to remark is the strong deviation of the intensity of
model 1DZ with respect to that in the other models, at all wavelengths
except the longest one.This strong deviation of the 1DZ
intensities is caused by the shallow temperature-tau relation shown in
Figure~\ref{fig:T_histogram_tau}, and is also clearly demonstrated in the
results of \citet{TrujilloBueno+Shchukina2009}.
It makes obvious that models derived by geometrically
averaging atmospheric properties are very far from reality for quantitative
spectroscopic analysis, except perhaps for very small $\mu$ values,
where the intensity forms in layers where the temperature and density
variations are strongly reduced.
At the shortest wavelength the averaged intensity from model MHD30G is
best approximated by 1DHSE, while the other equal-$\tau$ averaged
models underestimate the intensity, in particular at disk center.
However, at longer wavelengths the two $\tau$-averaged models perform
better. Clearly, none of the four \onedimensional\ models predicts the
CLV of continuum intensity accurately for the whole wavelength range. 
%

The differences between the intensities from the MHD30G (solid)
and 1DTAU (dashed) and 1DT4 (dot-dashed) models decrease with longer wavelengths,
vanishing almost completely at 4.7 $\mu$m, and are mainly the result of the 
non-linearity of the Planck function as function of temperature.
In LTE the emergent intensity is to first order given by the Planck function
at optical depth unity at the wavelength under consideration.
The surface-averaged intensity, of a \threedimensional\ atmosphere
is therefore given by:
\begin{equation}
  \left< I_{\lambda} \right>_{xy} = \left< B_{\lambda}(T) \right>_{\tau_{\lambda} = 1}
  \label{eq:EB}
\end{equation} 
Let the $\tau_{\lambda} = 1$ surface have temperature perturbations $\Delta T$
around an average temperature $\overline{T}$, with $\Delta T / \overline{T} \ll 1$.
We want to compare the spatially averaged emergent intensity at wavelength
$\lambda$ with the emergent intensity of a \onedimensional\ atmosphere with
temperature stratification equal to the average temperature of the inhomogeneous
model on each of its constant optical depth surfaces, i.e. we would like
to estimate
\begin{equation}
  \left< \Delta B_{\lambda} \right> = 
     \case{1}{2} \left< t^2 \right>
     \left. \frac{\textrm{d}^2 B_{\lambda}(t)}{\textrm{d} t^2} \right|_{t=0} +
       \mathcal{O}(\left< t^3 \right>),
  \label{eq:deltaB}
\end{equation}
defining $t \equiv \Delta T / \overline{T}$, as before, and remembering that
$\left<t\right> = 0$.
It is straightforward (if somewhat tedious) to show that the second derivative
of the Planck function is strictly positive with a value that strongly decreases
with wavelength (see Equation~\ref{eq:d2Bdt2}).
As a consequence the average radiation in the continuum coming from an
inhomogeneous atmosphere is always larger than that coming from a
homogeneous atmosphere with the same thermal stratification (average
temperature--tau relation),  consistent with the behavior of the differences 
between the emergent continuum intensities of models MHD30G and 1DTAU
as plotted in Figure~\ref{fig_int_CLV}.
To verify the accuracy of Equation~\ref{eq:deltaB} we calculated the values for
$\overline{T}$, $\left< \Delta T/\overline{T} \right>$,
and $\left. B''(t)\right|_{t=0}$ on the
surface of optical depth unity at 500 nm in model MHD30G, and found that
this estimate accounts for 86\% of the difference in disk-center intensity
with model 1DTAU at that wavelength. The rest of the difference can be ascribed
to inaccuracy of the Eddington-Barbier approximation
(Appendix~\ref{app:Planck}, Equation~\ref{eq:EB}),
and neglected higher orders in Equation~\ref{eq:deltaB}.

Because of the strong increase of the H$^-$ free-free opacity with wavelength,
the continuum in the infrared at 4.7 $\mu$m forms in higher atmospheric layers,
where the temperature difference between the models vanishes. Moreover,
at this long wavelength the
Planck function varies linearly with temperature (its second derivative vanishes),
thus reducing the effect of temperature inhomogeneities discussed above.
Both together cause the near equality of the 4.7 $\mu$m continuum curves in
Figure~\ref{fig_int_CLV}.

We thus confirm and explain the difference in CLV of continuum intensities
between one- and \threedimensional\ atmospheres with the same temperature
stratification as described by \citet{Koesterke+AllendePrieto+Lambert2008},
though a direct comparison is difficult since these authors compare their
results as a difference to observed CLV, and normalize to disk-center
intensities.
If we would interpret the observed CLV of intensity in the continuum, in particular
at shorter wavelengths, in terms of a \onedimensional\ atmosphere, we would
overestimate the average temperature gradient of the atmosphere (as measured
on surfaces of constant optical depth), since temperature inhomogeneities
raise the average emergent intensity at disk center, steepening the CLV.

Model 1DHSE is clearly too hot in the higher layers (see
Figure~\ref{fig:T_histogram_tau}), giving rise to an overestimate of
intensities, in particular at longer wavelengths, which form relatively
higher in the atmosphere.

\begin{figure}[bpth]
\epsscale{0.53}
\plotone{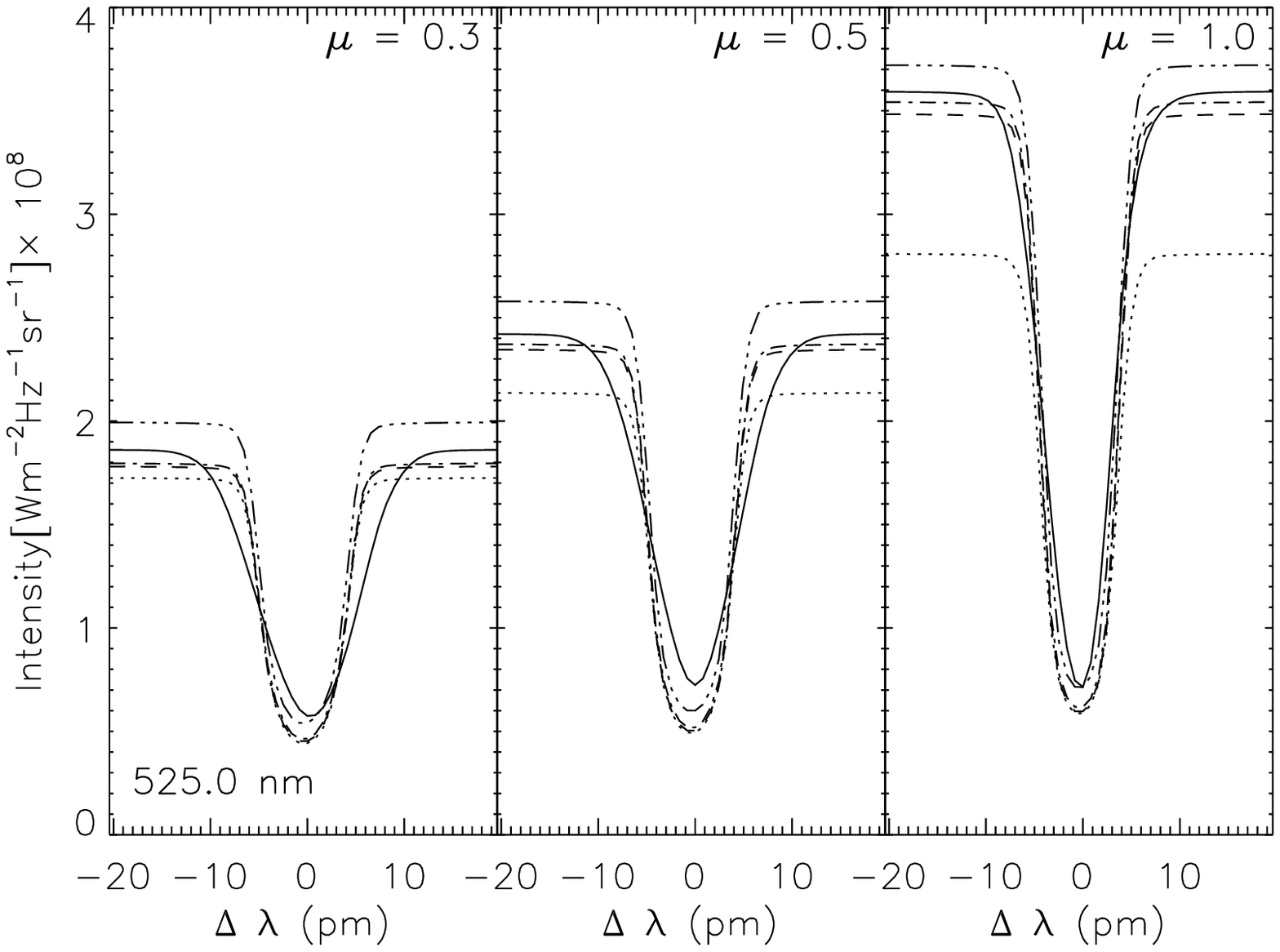}\\
\plotone{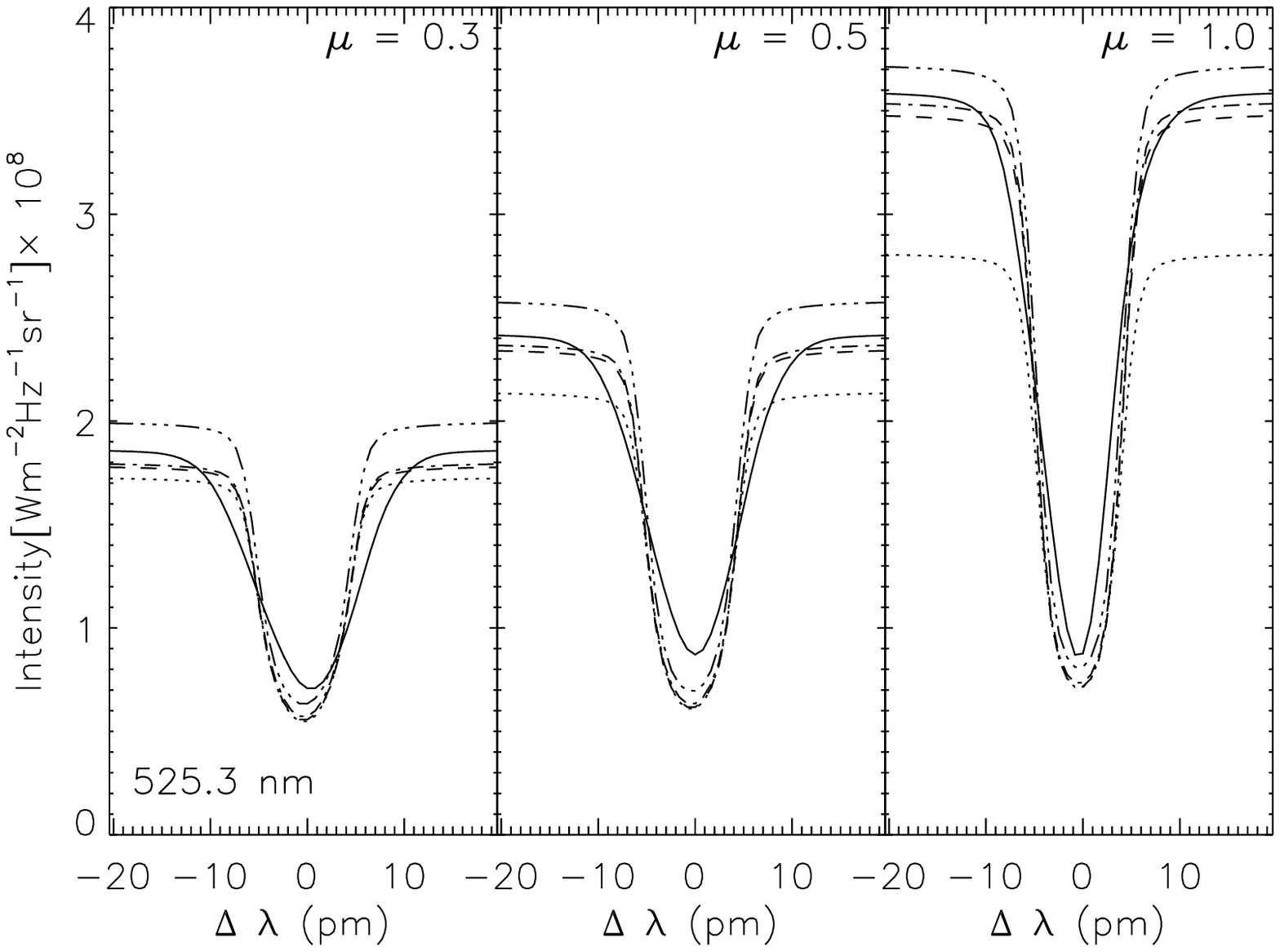}\\
\plotone{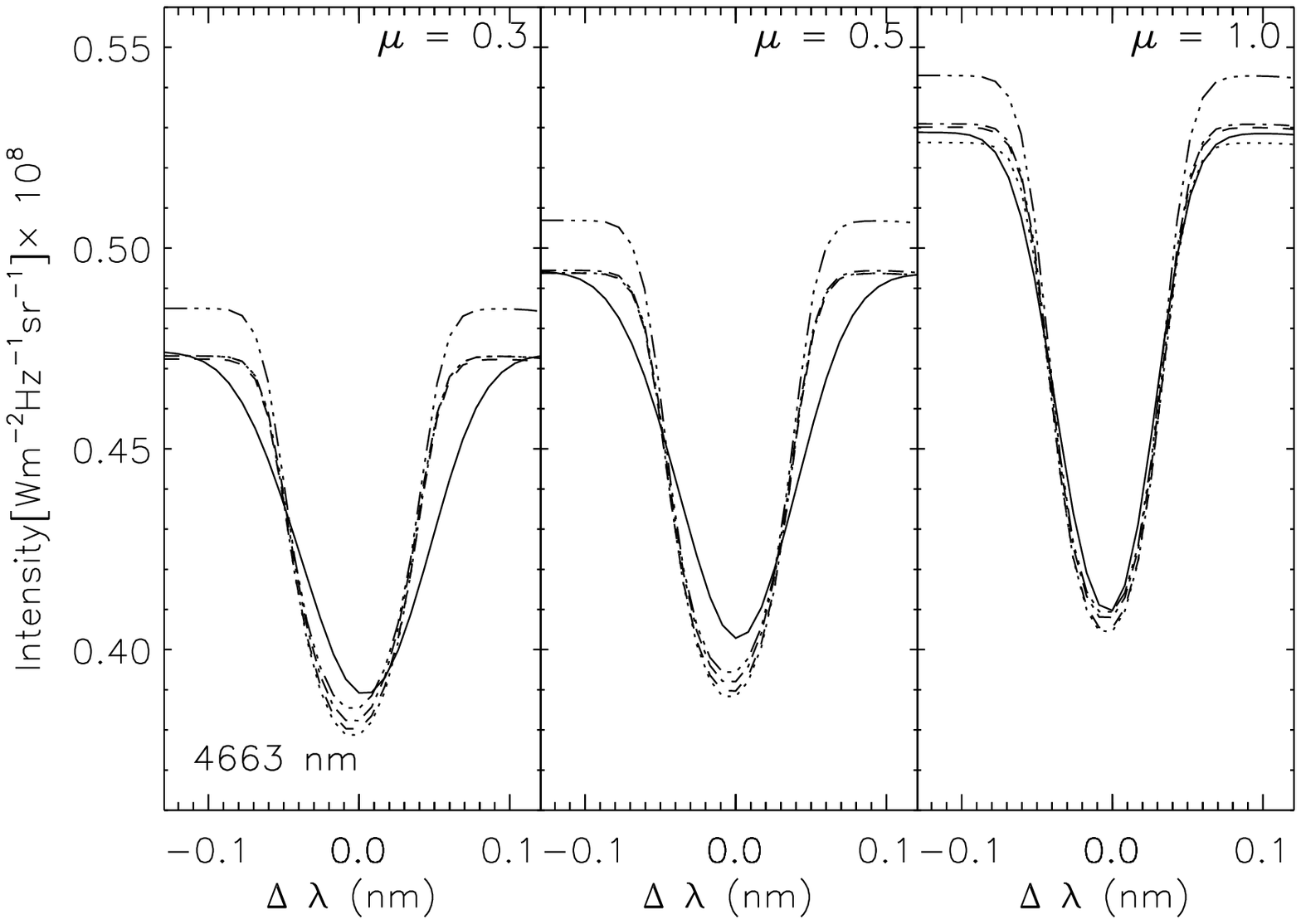}
\caption{Average line profiles for the MHD30G (solid curves), and
 1DTAU (dashed) and 1DZ (dotted), 1DT4 (dot-dashed) and 1DHSE (dot-dot-dot-dashed)
 models, for the viewing angles coresponding to $\mu = 0.3, 0.5$
 and $1.0$.\label{fig:lineprofs}}
\end{figure}
Calculated profiles of the \FeI\ 525.0 and 525.3 nm line and CO 7-6 R69 line
for three different viewing angles are shown in Figure~\ref{fig:lineprofs}.
As expected, the profiles from the \onedimensional\ models lack
the asymmetries produced by MHD30G.
In addition, we note that differences in intensity of the line-cores are smaller 
than in nearby continua, because the former originate higher in the atmosphere,
where amplitudes of temperature inhomogeneities are reduced (Figures~\ref{fig:T_histogram_z}
and \ref{fig:T_histogram_tau}).
However, the averaged line-core intensity in the MHD30G model is
systematically higher because of these remaining inhomogeneities, while the
core intensities of the two \onedimensional\ models are virtually
indistinguishable because they have the same temperature in the
core-forming part of the models, apart from model 1DHSE, which is too
hot in this region.

To estimate the effect the spatial averaging in the different models has on
the determination of element abundances we have matched the equivalent
widths in each of the \onedimensional\ models and lines with the one from
the average spectrum of model MHD30G by varying the abundances of iron
and oxygen, respectively.
The resulting differences are expressed in dex and tabulated in
Table~\ref{tab:abundances}, with negative values meaning
that the \onedimensional\ models would require a downward revision of
the abundance (and thus that their use would lead to an underestimate
of the abundance). Typical abundance differences are small but significant,
in agreement with the findings of \citet{Steffen+Holweger2002}, who
tabulate abundance determination differences between their \twodimensional\
convection simulation and similarly derived \onedimensional\ averaged models.
\begin{table}
\caption{Differences with abundance derived by matching
 the equivalent width of disk-center line profiles with that in MHD30G.\label{tab:abundances}}
\begin{tabular}{llll}
  \hline\hline
   $\lambda$ (nm) & 525.0  & 525.3 & 4663\\
  \hline
  1DZ & +0.06 & +0.09 & -0.02\\
  1DTAU & -0.09 & -0.09 & -0.07\\
  HSE & +0.03 & -0.02 & -0.15\\
  T4 & -0.07 & -0.09 & -0.03\\  \hline
\end{tabular}
\end{table}
We note that these determined abundance differences should in no way be
considered abundance corrections, as we have made no effort to properly
adjust microturbulence, and our estimates are based on the equivalent
width of disk center intensity rather than that of the disk-integrated flux.
Instead the numbers in Table~\ref{tab:abundances} are an indication of the
typical error that can be made by using different models.
Clearly, the abundance correction that would be implied by our CO line
analysis is opposite of what we would expect from the CO number density
that we presented in Section~\ref{sec:moldensity}. The bottom panel in
Figure~\ref{fig:lineprofs} makes evident what the reason for this discrepancy
is. In particular, the 1DHSE model has a significantly increased continuum level,
which increases the concomitant equivalent width, requiring a negative abundance
correction, rather than a positive. The temperature stratification of the 1DHSE
model here plays a larger role than the reduced CO concentration in that
stratification.

The CO line from MHD30G broaden its wings towards the limb (bottom panel
of Figure~\ref{fig:lineprofs}) because of the effect of horizontal convective
motions on the relatively narrow line, which has reduced thermal broadening
due to the weight of the CO molecule. This broadening would contribute
significantly to the equivalent width of the disk-integrated flux profile,
and would change the abundance correction we give in table~\ref{tab:abundances}
a more positive value, if we were to use the equivalent width of the flux profile rather
than that of the disk-center intensity. It is clear, therefore, that different
line formation effects can play opposite roles in the abundance determination.
These effects are very hard to keep track of in the gross simplifications
that are performed when representing a \threedimensional\ structured
atmosphere with a \onedimensional\ averaged model, no matter how sophisticated
the latter is constructed.

\begin{figure}[htbp]
\plotone{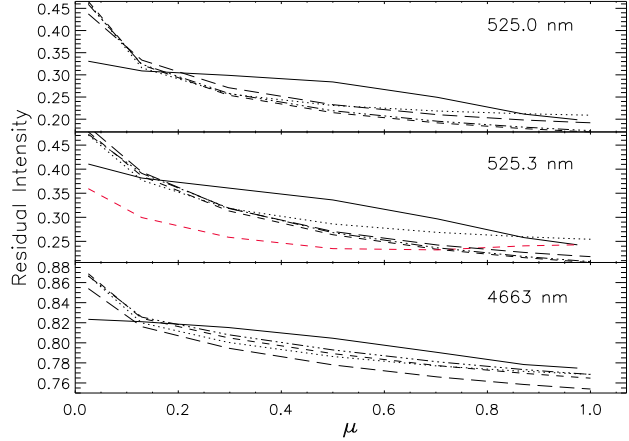}
\caption{CLVs of the residual line core intensity of the \ion{Fe}{1} 525.0,
  525.3 nm and 4663.6 nm CO lines in the five models MHD30G (solid curves), 1DTAU
  (dashed), 1DZ (dotted), 1DT4 (dot-dashed) and 1DHSE (dot-dot-dot-dashed).
  Red dashed curve shows the average resudual line core intensity behavior
  in the \threedimensional\ model when horizontal velocities are set to zero.\label{fig:linecores}}
\end{figure}
%
Differences between line spectra in the one- and \threedimensional\ models
are even more evident in the CLV of the residual line-core intensities
(defined as $(I_{\textrm{cont}} - I_{\textrm{core}}) / I_{\textrm{cont}}$) plotted
in Figure~\ref{fig:linecores}.
Strikingly, the residual line-core intensities of the two iron lines in model MHD30G 
show a much shallower decrease with decreasing $\mu$ than the core intensities
in the averaged models. Given the imposed equality
of the average thermal stratifications of models 1DTAU and MHD30G, the stark
difference between their line-core CLVs must be the result of mass motions that
are properly taken into account in the \threedimensional\ model, but represented
by microturbulence in the \onedimensional\ derivatives. Indeed,
by eliminating the horizontal and vertical velocities, respectively, we determined
that the CLV behavior of the line cores is the result of the presence of horizontal
motions in the convective model. Setting these to zero results in a CLV that 
behaves similarly as function of $\mu$ as in the \onedimensional models
(red dashed line in Figure~\ref{fig:linecores} in the midle panel).

The CLV behavior of the average residual line-core intensity in the MHD30G
model can be explained in the following way.  For moderate viewing
angles the horizontal convective velocities shift the core opacity out
of the center of the line, lowering the formation height at those
wavelengths, sampling higher temperatures, broadening the line and
increasing the line core intensity.  By contrast, the vertical
velocities result in the familiar convective blue-shift because of the
combined effect of the area asymmetries of down- and up-flows and the
correlation between brightness and up-flows.  Clearly, the line-core
CLV will affect disk-integrated line intensities, and other derived
line parameters, which are used in stellar abundance determinations.

\section{Discussion and conclusion\label{sec:conclusions}}
The analysis of stellar spectra would greatly benefit if it would be
possible to extract many physical parameters accurately through
\onedimensional\ modeling. The radiative transfer in this type
of models is orders of magnitude less computationally demanding
than in self consistent and more realistic \threedimensional\
(M)HD models.
In this paper we have attempted to further clarify if such a less
numerically demanding approach is feasible, and if not, what the
physical reasons for failure could be. We have approached this
question in the simplest possible fashion by comparing spectra
in several continua and lines from \onedimensional\ atmospheres
that were derived from a \threedimensional\ MHD snapshot
by straight averaging over equal geometric heights and
over surfaces of equal optical depth at 500 nm.
In this way we assured as much as possible that the derived \onedimensional\
atmospheres had the same average properties as the original snapshot, 
so that possible differences between the spectra are mostly the
result of the inhomogeneities in the \threedimensional\ model.
As a result we have identified several mechanisms that affect the
average spectrum of the inhomogeneous \threedimensional\ model that
cannot be adequately represented in derived \onedimensional\ models,
and thus would lead to wrong estimates of the physical parameters extracted from
the spectrum, if that spectrum were to be interpreted in the context of
simpler \onedimensional\ models.

In particular, we constructed two types of \onedimensional\ spatially averaged atmospheric
stratifications,
1DZ, and 1DTAU and 1DT4, respectively, by averaging the thermodynamic
properties of a single snapshot, MHD30G,
from simulation of solar magneto-convection, over surfaces of equal
geometric height and equal optical depth at 500 nm.
In addition, we created a model, 1DHSE, by taking model 1DTAU and allowing its
stratification to be determined by hydrostatic equilibrium.
We then calculated the intensities in continua at 400, 500, 800 nm, and 4.7 $\mu$m,
two \ion{Fe}{1} lines at 525.0 and 525.3 nm,
and a CO molecular line at 4663 nm from all five models,
compared the spatially averaged spectra from the \threedimensional\ model
at different viewing angles with those from the average-property \onedimensional\
models at the same angles.

From the calculated spectra it is immediately clear that the geometrically averaged
model 1DZ provides a particularly poor representation of the average properties
of the \threedimensional\ model in terms of spectroscopic diagnostics.
The geometric averaging results in a temperature--$\tau$ stratification that is much
shallower than the stratification of average temperature
in the inhomogeneous atmosphere, and thus results in much lower continuum intensities
at disk center, and a much shallower CLV.
Only very close to the limb where intensities emanate from relatively higher layers of the
atmosphere that are less affected by granulation and thus more homogeneous,
and where the obliquity of the rays naturally averages over horizontal inhomogeneities,
the intensities from this model converge on the averaged ones from MHD30G.

It is no surprise that intensities derived from models 1DTAU and 1DT4 are much closer to the
spatially averaged ones from model MHD30G, as radiation at a given wavelength
naturally comes from a layer with limited range in optical depth around unity.
Yet, even in continua the average intensity from model MHD30G lies well above
those of the tau-averaged models, in particular at shorter wavelengths, and closer to disk
center. This underestimate of continuum intensities by the \onedimensional\ models
is the result of the non-linear mapping of temperature inhomogeneities into
intensity fluctuations by the Planck function. In Section~\ref{sec:spectra}
we show that the average intensity
of an atmosphere with temperature fluctuations is always higher than the
intensity of a homogeneous atmosphere with the same average temperature stratification,
because of the strict positivity of the second derivative of the Planck function
with temperature. For large temperature and/or long wavelengths the second
derivative of $B_{\lambda}$ vanishes, and temperature inhomogeneities are mapped
linearly into intensity variations, so that differences between the \threedimensional\
and derived \onedimensional\ models disappear, as is evident in the continuum
at 4.7 $\mu$m (Figure~\ref{fig_int_CLV}).
A corollary of the non-linear mapping of the Planck function is that the interpretation
of an average spectrum from a horizontally inhomogeneous atmosphere, in terms of
a one-dimensional model will lead to an overestimate of the average temperature
gradient in the atmosphere.

Molecular lines are often used in abundance determinations, because they are generally
weak, less affected by thermal broadening because of their mass, and plentiful.
However, their concentration is sensitive to thermal conditions. To investigate
how temperature and density inhomogeneities affect the average concentration
molecules and emergent spectra of their lines, we calculated the CO number
densities in the inhomogeneous model MHD30D and averaged models 1DTAU and 1DZ,
as well as the emergent spectrum of the CO 7-6 R68 rotation-vibration line for
different viewing angles. In Section~\ref{sec:moldensity} we show that the presence of both
temperature and density inhomogeneities to first order always lead to an
\textit{increase} of the average molecular density of a diatomic molecule over the
values in a homogeneous atmosphere with the same average temperature and density
stratification. As an example we show the calculated number density of the
CO molecule in models MHD30G and 1DTAU as a function of average optical depth in
Figure~\ref{fig:COnum}.
The underestimate of molecular densities by a one-dimensional model that is used
to represent the average of an inhomogeneous atmosphere will lead to an overestimate
of the abundances of the constituent atoms of the molecule if molecular lines
are used for determination of these abundances.

It is obvious that \onedimensional\ average-property models cannot represent
the complex mass motions that result from convection. One effect that the 
average models necessarily fail to incorporate is the convective blue shift
and line characteristic ``C-shaped'' asymmetry of the bisectors of
spectral lines that results from the correlation of brightness and upflows and
asymmetry between up- and downflows in the convective flows.
We identify one other important aspect of the convective flows that \onedimensional\
models lack, namely the broadening and line weakening that horizontal motions
produce for moderate viewing angles ($\mu \approx 0.7$). The weakening results
from the shift of opacity out of the line core by horizontal motions, both
to the red and to the blue, increasing formation height and raising the central intensity.
This effect leads to an initial increase in the line-core intensity as function of
increasing viewing angle (decreasing $\mu$), and a much shallower decline of the
residual line-core intensity as is clearly shown in Figure~\ref{fig:linecores}.
The horizontal velocity effect is apertly absent in the CLVs obtained from the
\onedimensional\ models.
It thus also affects the disk-integrated line profile shape that will therefore be
wrongly interpreted if analyzed in the context of such average-property models.
Observational evidence for the the shallow behavior of the residual line-core
intensity is seen in \citet{RodriguezHidalgo+Collados+Vazques1994} who determined
the CLV of this property in the \ion{Fe}{1} 593.02 nm line, among others.
Curiously, the \ion{Mn}{1}  539.47 nm line shows behavior more in line with our
results from the \onedimensional\ models. This is perhaps the result of the
large hyperfine structure broadening of manganese lines, which makes them
less susceptible to convective velocity broadening, as explained by
\citet{Vitas+Viticchie+Rutten+Voegler2009}.

In summary, our comparison of spectra, and analysis of the physical effects of
temperature and density inhomogeneities make clear that extreme caution has to
be taken when the average spectrum of a horizontally inhomogeneous atmosphere
is interpreted in the context of a \onedimensional\ atmospheric representation.
Not only the lack of incorporation of convective motions into \onedimensional\
models, but also the non-linearities involved in establishing molecular
equilibrium and level populations as function of temperature and density,
and the non-linearities of the Planck function as function of temperature
likely lead to misinterpretation of the observed spectrum.
In the work here we have concentrated on spatial averaging, but our derivations
and contentions hold equally well in case of temporal averaging.
In addition, we suspect that non-linearities introduced by non-LTE conditions,
which we have not addressed in this paper, have to be treated with even more
caution. The realism of chromospheric modeling, which still relies heavily on
one-dimensional models, therefore needs careful examination in this respect.
A very illustrative example is given by the simulations
presented by \citet{Carlsson+Stein1994}, who simulate acoustic shock propagation
in the solar atmosphere. The shocks heat the atmosphere, but their main effect
is an enhancement of chromospheric emission. When the resulting time-averaged
spectrum is modeled with a \onedimensional\ hydrostatic semi-empirical model,
the resulting atmosphere is required to have a chromospheric temperature rise.
This temperature rise is very much at odds with the mean temperature stratification
of the simulations which shows a monotonic decline with height.

Since the 1DTAU model we employed exactly matches the average thermal stratification
of the \threedimensional\ snapshot, but produces different intensity values
in lines, and even continua, we conclude that a one-dimensional model that
reproduces the average spectrum of an inhomogeneous atmosphere necessarily
has a different average stratification than this atmosphere. It is therefore
in our opinion not possible to produce a one-dimensional atmospheric model that is at
the \textit{same time} spectroscopically equivalent \textit{and} has matching
average physical properties.


\bibliographystyle{aastex5} \bibliography{uitenbr}

\begin{appendix}

\section{Second derivative of $\phi$\label{app:phi}}
The function $\phi(T)$ is defined (Equation~\ref{eq:nAB}) as:
\begin{equation}
  \phi(T) \equiv \left(\frac{D}{kT} \right)^{3/2} \exp{\left[\frac{D}{kT}\right]}.
\end{equation}
Writing $T = \overline{T} + \Delta T$ and defining $t \equiv \Delta T / \overline{T}$
we can write $\phi$ as a function of $t$:
\begin{equation}
  \phi(t) = \left(\frac{D}{k \overline{T}} \right)^{3/2} \frac{1}{(1 + t)^{3/2}}
             \exp{\left[\frac{D}{k \overline{T}} \left( \frac{1}{1 + t} \right)\right]}.
\end{equation}
It follows that the first derivative of this function with respect to $t$ is
\begin{equation}
  \phi'(t) = - \left\{ \frac{3}{2} \frac{1}{(1+t)} + \frac{D}{k \overline{T}} \frac{1}{(1+t)^2}\right\}
               \phi(t),
\end{equation}
and its second derivative
\begin{equation}
  \phi''(t) = \left\{\frac{15}{4} \frac{1}{(1+t)^2} + 5 \frac{D}{k \overline{T}} \frac{1}{(1+t)^3} +
                \left( \frac{D}{k \overline{T}} \right)^2  \frac{1}{(1+t)^4} \right\} \phi(t).
\end{equation}

\section{Second derivative of the Planck function\label{app:Planck}}
The Planck function for wavelength $\lambda$ and temperature $T$ is given by:
\begin{equation}
  B_{\lambda}(T) = \frac{2 h c^2}{\lambda^5} \frac{1}{e^{\frac{h c}{\lambda k T}} - 1}
\end{equation}
Writing $T = \overline{T} + \Delta T$, and defining $t \equiv \Delta T / \overline{T}$ and 
$b \equiv hc/ \lambda k \overline{T}$, we rewrite $B$ as a function of $t$:
\begin{equation}
  B_{\lambda}(t) = \frac{2 h c^2}{\lambda^5} \frac{1}{e^{\frac{b}{1 + t}} - 1}
\end{equation}
It follows that 
\begin{equation}
  B'_{\lambda}(t) = \frac{b}{(1 + t)^2} \frac{1}{1 - e^{\frac{-b}{1+t}}} B_{\lambda}(t),
\end{equation}
and
\begin{equation}
  B''_{\lambda}(t) = \frac{b}{1 - e^{\frac{-b}{1+t}}} \frac{1}{(1+t)^3} \left\{
                      \left(\frac{1 + e^{\frac{-b}{1+t}}}{1 - e^{\frac{-b}{1+t}}} \right) \frac{b}{1+t} \right\}
                       B_{\lambda}(t).
\end{equation}
For $t = 0$ the second derivative is
\begin{equation}
  \left. B''_{\lambda}(t)\right|_{t=0} = \frac{b}{1 - e^{-b}} \left\{ \left(\frac{1 + e^{-b}}{1 - e^{-b}} \right) b -
                                         2\right\} \left. B_{\lambda}(t)\right|_{t=0}.
  \label{eq:d2Bdt2}
\end{equation}
The expression between curly brackets in Equation~\ref{eq:d2Bdt2} is strictly positive for all $b$,
and vanishes for $b << 1$, i.e., for large wavelength, large temperature, or both.

\end{appendix}

\end{document}